\documentstyle[prd,aps,psfig]{revtex}

\begin{document}
\large \draft
\title{Nonsingular FRW cosmology and nonlinear electrodynamics}
\vspace{0.5cm}
\author{C. S. C\^amara$^{1}$\thanks{Electronic address: calist@dfte.ufrn.br},
 M. R. de Garcia Maia$^{1}$\thanks{Electronic address: mrgm@dfte.ufrn.br},
J. C. Carvalho$^{1}$\thanks{Electronic address:
carvalho@dfte.ufrn.br}, J. A. S. Lima$^{1,2}$\thanks{Electronic
address: limajas@dfte.ufrn.br, limajas@astro.iag.usp.br}}
\address{$^1$Departamento de F\'{\i}sica, Universidade Federal do Rio
Grande do Norte, 59072-970 Natal RN Brazil\\ $^{2}$Instituto de
Astronomia, Geof\'{\i}sica e Ci\^encias Atmosf\'ericas,
Universidade de S\~ao Paulo\\ Rua do Mat\~ao, 1226 - Cidade
Universit\'aria, 05508-900, S\~ao Paulo, SP, Brasil}
\maketitle


\vspace{1.0cm}

\begin{abstract}
\large
The possibility to avoid the cosmic initial singularity as a
consequence of nonlinear effects on the Maxwell eletromagnetic
theory is discussed. For a flat FRW geometry we derive the general
nonsingular solution supported by a magnetic field plus a cosmic
fluid and a nonvanishing vacuum energy density. The nonsingular
behavior of solutions with a time-dependent $\Lambda(t)$-term are
also examined. As a general result, it is found that the
functional dependence of $\Lambda(t)$ can uniquely be determined
only if the magnetic field remains constant. All these models are
examples of bouncing universes which may exhibit an inflationary
dynamics driven by the nonlinear corrections of the magnetic
field.
\end{abstract}
\pacs{PACS number(s): 98.80.Bp, 11.10.Lm.40.Nr 98.80.Cq}
\section{Introduction}

\label{s1}

A fundamental difficulty underlying the standard
Friedmann-Robertson-Walker (FRW) cosmology is the prediction of an
initial singular state where all curvature invariants and some
material quantities like pressure, energy density and temperature
become infinite\cite{Wheeler}. Generically, ``the break down of
the laws of physics at a singularity" is a clear manifestation of
mathematical inconsistency and physical incompleteness of any
cosmological model. In this way, although strongly supported by
the recent observations at low and intermediate redshifts, the
present big-bang picture with dark energy must be improved at very
high redshifts.

In order to solve such a problem several attempts based on many
disparate mechanisms have been proposed in the literature. Some
earlier approaches trying to develop a well behaved and more
complete cosmological description include: quadratic Lagrangians
and other alternatives theories for the gravitational
field\cite{QL}, a creation-field cosmology\cite{pad}, a huge
vacuum energy density at very early times\cite{VED}, nonminimal
couplings\cite{NMC}, nonequilibrium thermodynamic
effects\cite{NET}, and quantum-gravitational phenomena closely
related to a possible spontaneous birth of the
universe\cite{Kandrup}.

More recently, a new interesting mechanism aiming to avoid the
cosmic singularity has been discussed by De Lorenci et
al.\cite{novello98a} through a nonlinear extension of the Maxwell
eletromagnetic theory. The associated Lagrangian and the resulting
electrodynamics can theoretically be justified with basis on
different arguments. For example, the nonlinear terms can be added
to the standard Maxwell Lagrangian by imposing the existence of
symmetries such as parity conservation, gauge invariance, Lorentz
invariance, etc\cite{novello96,Munoz96}, as well as by the
introduction of first order quantum corrections to the Maxwell
electrodynamics\cite{heisenberg36,schwinger51}. It is worth notice
that nonlinear corrections may also be important to avoid the
black hole singularity. Actually, an exact regular black hole
solution has been recently obtained with basis on the
Einstein-dual nonlinear electrodynamics as proposed by Salazar,
Garcia and Plebansky \cite{SP87,BG98}. Note, however, that the
purpose of the present work is not to give a detailed description
of the possible nonlinear theories and their physical effects. Our
basic aim is rather modest - we will try to gain new insights into
the possibilities beyond Maxwell theory based on the simplest
nonlinear electromagnetic lagrangian and its connection to the
cosmic singularity problem.

In this concern, it has been found \cite{novello98a} that the
primordial singularity can be removed because the nonlinear
corrections reinforce the negative pressure at the early stages of
the universe. The authors also argued that the nonsingular
behaviour is unaffected by the presence of other ultrarelativistic
components obeying the equation of state $p_{(ur)} = \rho_{(ur)}
/3$.

In the present work, we extend the analysis by De Lorenci et
al.\cite{novello98a} using different ingredients. We first obtain
the behavior of the scale function for a spatially flat FRW
geometry ($\Lambda=0$), thereby showing that their results is a
particular case of the general solution. It is also analyzed how
this solution is modified by the presence of a $\Lambda$-term
(vacuum energy density) which may be constant or a time varying
quantity. Solutions with constant vacuum energy density also leads
to a non-singular universe, and the same happens with a
time-dependent $\Lambda(t)$-term. However, there are singular
solutions where the decaying vacuum supplies the energy to a
constant cosmological magnetic field. In this case, the time
dependence of the cosmological term can uniquely be determined and
corresponds to a slightly modification of the more frequent forms
suggested in the literature\cite{OT,CLW92,overduin98}. The main
physical restrictions, including the time interval where the
nonlinear corrections must be important are also discussed.

The article is organized as follows. In section \ref{s2}, we write
down the Einstein field equations (EFE) for a flat FRW geometry
supported by an energy momentum derived from the extended
(nonlinear) eletromagnetic Lagrangian. In section \ref{s3}, we
generalize the solution derived in \cite{novello98a} which assumes
a vanishing cosmological term, a spatially flat geometry and a
time-dependent magnetic field. In section \ref{s4}, we obtain a
new solution that takes into account the presence of a constant
$\Lambda$. The behavior involving a time-dependent $\Lambda$ is
discussed in section \ref{s5}. Finally, in the conclusion section,
we summarize the basic results and also present some suggestions
for future work. In what follows, Greek indices run from 0 to 3,
Latin indices run from 1 to 3. Unlike of Ref.\cite{novello98a}, we
adopt the International System of Units (see the Appendix on units
and dimensions of Ref. \cite{jackson} for further details.)

\section{Basic Equations}
\label{s2}
As widely known, the Lagrangian density for free fields in the
Maxwell electrodynamics may be written as
\begin{equation}
\label{e1} \protect{\cal{L}}_{(MAXWELL)}
=-\frac{1}{4\mu_0}F^{\mu\nu}F_{\mu\nu} = -\frac{1}{4\mu_0}F\;,
\end{equation}
where $F^{\mu\nu}$ is the electromagnetic field strength tensor
and $\mu_0$ is the magnetic permeability. The canonical
energy-momentum tensor is then given by
\begin{equation}
\label{e2} T_{\mu\nu}^{(MAXWELL)}
=\left(\frac{1}{\mu_0}\right)\left[
F_{\mu\alpha}{F^{\alpha}}_{\nu}+\frac{1}{4}Fg_{\mu\nu} \right].
\end{equation}

In this work we consider the extended Lagrangian density to the
electromagnetic field
\begin{equation}
\label{e3} \protect{\cal{L}}=-\frac{1}{4\mu_0}F +\omega F^2 +
\eta{F^{*}}^2,
\end{equation}
where $\omega$ and $\eta$ are arbitrary constants,
\begin{equation}
\label{e4} F^{*}\equiv F^{*}_{\mu\nu}F^{\mu\nu}\;,
\end{equation}
and $F^{*}_{\mu\nu}$ is the dual of $F_{\mu\nu}$. As one may
check, the corresponding energy-momentum tensor becomes
\begin{equation}
\label{e5} T_{\mu\nu}=-4\frac{\partial\protect{\cal{L}}}{\partial
F}{F_{\mu}}^{\alpha}F_{\alpha\nu}+\left
(\frac{\partial\protect{\cal{L}}}{\partial
F^{*}}F^{*}-\protect{\cal{L}}\right )\,g_{\mu\nu}\;.
\end{equation}
Let us now consider the above expressions in the context of a
homogeneous and isotropic FRW flat model
\begin{equation}
\label{e6} ds^2 = c^2dt^2-a^2(t)\left
[dr^2+r^2d\theta^2+r^2\sin^2\theta \,d\phi^2\right]\;.
\end{equation}
Naturally, electromagnetic fields may source of the above
background only if the fields are considered in its average
properties\cite{tolman30}. Now, applying the standard spatial
averaging process we set
\begin{eqnarray}
\label{e7} <E_i> &=& 0\;,\\ \label{e8} <B_i>&=& 0\;,\\ \label{e9}
<E_iE_j>& =& -\frac{1}{3} E^2 g_{ij}\;,\\ \label{e10} <B_i
B_j>&=&-\frac{1}{3} B^2 g_{ij}\;,\\ \label{e11} <E_iB_j>&=& 0\;.
\end{eqnarray}
Equations (\ref{e7}) - (\ref{e11}) imply that
\begin{equation}
\label{e12} <F_{\mu\alpha}{F^{\alpha}}_{\mu}>=\frac{2}{3}
\left(\epsilon_0E^2+\frac{B^2}{\mu_0}\right)
\frac{U_{\mu}U_{\nu}}{c^2}+\frac{1}{3}
\left(\epsilon_0E^2-\frac{B^2}{\mu_0}\right)g_{\mu\nu}\,,
\end{equation}
where $U_{\mu}$ is the four velocity. Under such conditions the
average value of the energy-momentum tensor takes the perfect
fluid form, namely:
\begin{equation}
\label{e14}
<T_{\mu\nu}>=(\rho+p)\frac{U_{\mu}U_{\nu}}{c^2}-pg_{\mu\nu}
\end{equation}
where the density $\rho$ and pressure $p$ have the well known form
\begin{eqnarray}
\label{e15} \rho&=&
\frac{1}{2}\left(\epsilon_0E^2+\frac{B^2}{\mu_0}\right),\\
\label{e16} p&=& \frac{1}{3}\; \rho\;.
\end{eqnarray}

In order to analyze the modifications implied by the use of the
modified Lagrangian (\ref{e3}), we also assume that for the
stochastically defined electromagnetic fields only the average
value of the squared magnetic field $B^2$ survives at the very
early Universe, i. e., we use Eqs. (\ref{e7}) - (\ref{e11}) with
$<E^2>=0$. In this concern, we remark that a homogeneous electric
field in a plasma must give rise to an electric current of charged
particles and then rapidly decay. Indeed, unlike what happens with
the magnetic field, at present there is no basis whatsoever to
presume the existence of an overall electric field. Indeed, since
the late sixties, it has been recognized that cosmological models
with an overall electric field bears hardly any relation at all to
reality (see for instance, Zeldovich and Novikov
\cite{Zeldovich60}). Naturally, this does not means that the
$<E^{2}>$ term appearing in our stochastic approach can be
neglected in comparison with $<B^{2}>$, but one may expect that
its influence might be small for some special regimes, as for
example, when the plasma may be treated with basis on the
magnetohydrodynamics approximation. For a dense ionized gas, for
example, the collision frequency can be so high that the electric
field and its momenta may arise only as a consequence of the
motion of the fluid, or as a result of the external charges
distribution or (not ``frozen in") time-varying magnetic fields.
Such a behavior may happen in the primeval plasma (below Planck's
temperature) because the Debye screening radius $\sim (T/n)^{1/2}$
is very small in comparison with the macroscopic relevant scale
for nonsingular world models, namely, the Hubble radius
$cH(t)^{-1}$. Keeping these remarks in mind, we return to the
basic equations by assuming that $<E^{2}>$  has been neglected so
that (\ref{e14}) still holds, but the energy density and pressure
now read

\begin{eqnarray}
\label{e17} \rho&=&\frac{1}{2\mu_0}B^2\left(1-8\mu_0\omega
B^2\right),\\ \label{e18} p&=&\frac{1}{6\mu_0}
B^2\,(1-40\mu_0\omega B^2) = \frac{1}{3}\rho-\frac{16}{3}\omega
B^4\;.
\end{eqnarray}
Note that the weak energy condition ($\rho > 0$) is obeyed for
\begin{equation}
\label{e19} B < \frac{1}{2\sqrt{2\mu_0\omega}},
\end{equation}
whereas the pressure will reach negative values only if
\begin{equation}
\label{e20} B > \frac{1}{2\sqrt{10\mu_0\omega}}.
\end{equation}

On the other hand, there is a widespread belief that the early
Universe evolved through some phase transitions, thereby yielding
a vacuum energy density which at present is at least $118$ orders
of magnitude smaller than in the Planck time\cite{Weinberg}. Such
a discrepancy between theoretical expectation (from the modern
microscopic theory of particles and gravity) and empirical
observations constitutes a fundamental problem in the interface
uniting astrophysics, particle physics and cosmology, which is
often called ``the cosmological constant problem''
\cite{Weinberg,Dolgov}. This ``puzzle" together with the
observations of type Ia Supernovae\cite{perlmutter98} suggesting
that the cosmic bulk of energy is repulsive and appears like a
dark energy (probably of primordial origin) stimulated the recent
interest for more general models containing an extra component and
accounting for the present accelerated stage of the Universe
\cite{prp}. A possible class of such cosmologies is provided by
phenomenological models driven by a constant or a time-dependent
$\Lambda(t)$-term (see, for instance, [14-16, 24-30] and
references therein). The effective time-dependent cosmological
term may be regarded as a second fluid component with energy
density, $\rho_{\Lambda}(t) =\Lambda(t)c^4/{8 \pi G}$, which
transfers energy continuously to the material medium. The
conditions under which this kind of cosmology can be described by
a scalar field coupled to a perfect fluid has also been discussed
in the literature [24-30]. For the sake of generality, we focus
our attention to a decaying $\Lambda$ model, but now in the
presence of the primeval magnetic field. In the background defined
by (\ref{e6}), the EFE read
\begin{equation}
\label{e21} \frac{\dot{a}^2}{a^2}
=\frac{8\,\pi\,G}{3c^2}\,\rho+\frac{\Lambda(t)c^2}{3}\;,
\end{equation}
\begin{equation}
\label{e22} \frac{\ddot{a}}{a}=\frac{\Lambda(t)c^2}{3} -
\frac{4\,\pi\,G}{3c^2}\,(\rho+3p)\;,
\end{equation}
while the energy conservation law can be written as
\begin{equation}
\label{e23} \dot{\rho} +3\,\frac{\dot{a}}{a}\,(\rho
+p)=-\frac{\dot{\Lambda}c^4}{8\,\pi\,G}\;.
\end{equation}
where an overdot means derivative with respect to the cosmic time
$t$. As we shall see, for a constant magnetic field it is not need
to assume a phenomenological expression for $\Lambda(t)$ as
usually done since it can be uniquely fixed from the above set of
equations.

Replacing (\ref{e17}) and (\ref{e18}) in the Einstein equations
(\ref{e21}) - (\ref{e23}) we get
\begin{equation}
\label{e25}
\frac{\dot{a}^2}{a^2}=\frac{4\,\pi\,G}{3c^2}\,\frac{B^2}{\mu_0}\,(1-8\,\mu_0\,
\omega B^2)+\frac{\Lambda c^2}{3}\;,
\end{equation}
\begin{equation}
\label{e24} \frac{\ddot{a}}{a}=\frac{\Lambda c^2}{3} -
\frac{4\,\pi\,G}{3c^2}\,\frac{B^2}{\mu_0}\,(1-24\,\mu_0\, \omega B^2)\;,
\end{equation}
\begin{equation}
\label{e26} \frac{B}{\mu_0}\,\left (1-16\,\mu_0\,\omega B^2\right
)\,\left
(\dot{B}+2\,\frac{\dot{a}}{a}\,B\right)=-\frac{{\dot{\Lambda}
{c^4}}}{8\,\pi\,G}\;.
\end{equation}
Inserting equations (\ref{e17}) and (\ref{e21}) into (\ref{e22})
we find
\begin{equation}
\label{efe1}
a\ddot{a}+{\dot{a}}^2-\frac{2}{3}\Lambda c^2a^2- \left(\frac{64\pi
Gw}{3c^2}\right)B^4a^2=0.
\end{equation}
The term proportional to $B^{4}$ comes from the nonlinear
correction in the Lagrangian. Note also that if the coupling
constant $w$ is zero, the standard FRW differential equation for a
radiation filled universe plus a cosmological $\Lambda$-term is
recovered. By solving any two of the above equations one may
discuss if the nonlinear terms added to the Maxwell Lagrangian may
alter the primeval singular state. In order to compare with some
previous results presented in the literature, we start our
analysis by considering some interesting particular cases.
\section{Nonsingular models with $\Lambda=0$}
\label{s3}
This is the case studied by De Lorenci et al.\cite{novello98a}
where a particular nonsingular solution was found. In what follows
the corresponding general solution is given and we also show how
to recover the quoted result. First of all, we remark that if $B$
is a time-dependent quantity and $\Lambda$ remains constant,
equation (\ref{e26}) can be easily integrated to give
\begin{equation}
\label{e27} B(t) = B_0 \,\left (\frac{a_0}{a}\right )^2\;,
\end{equation}
where $B_0$ is a constant of integration. In this paper, the
subscript $0$ does {\em not} mean the present day value of a
quantity. It only indicates its value at an arbitrary time $t_0$
which appears in the general solution of $a(t)$ as a second
integration constant. This constant $t_0$ was arbitrarily chosen
in \cite{novello98a}. Thus $B_0 = B(t_0)$ if $a_0=a(t_0)$ (in the
quoted paper $a_0$ was also normalized to unit). We stress that
the scaling solution (\ref{e27}) holds even for a nonvanishing
constant $\Lambda$. Inserting $B(a)$ into (\ref{efe1}) one finds
\begin{equation}
\label{efe2}
a\ddot{a}+{\dot{a}}^2- \left(\frac{64\pi
GwB_0^4a_0^8}{3c^2}\right)a^{-6}=0.
\end{equation}
 As one may check, the general solution of the above equation is

\begin{equation}
\label{e28} a(t)= a_0\,\left [
4\,\alpha_0^2\,(t-t_0)^2\,+\,4\,\alpha_0\,\beta_0
\,(t-t_0)\,+\,1\right]^{1/4}\;,
\end{equation}
where we have defined
\begin{equation}
\label{e29} \alpha_0\equiv \sqrt{\frac{4\,\pi\,G}{3\mu_0c^2}}\,B_0\;,
\end{equation}
\begin{equation}
\label{e30} \beta_0\equiv \pm\sqrt{1-8\,\mu_0\, \omega B_0^2}\;.
\end{equation}

In order to compare with the results of \cite{novello98a} we
recast (\ref{e28}) in the form
\begin{equation}
\label{e31}
a(t)=a_0\,(4\,\alpha_0^2\,t^2\,+\,4\,\alpha_0\,\gamma_0\,t\,+\,\delta_0)^{1/4}\;,
\end{equation}
with
\begin{equation}
\label{e32} \gamma_0\equiv\beta_0-2\alpha_0t_0\;,
\end{equation}
\begin{equation}
\label{e33} \delta_0\equiv
4\alpha_0t_0\,(\alpha_0t_0-\beta_0)+1\;.
\end{equation}
The linear term in $t$ inside the parenthesis of (\ref{e31}) does
not appear in the solution given by the authors of Ref.
\cite{novello98a}. This happens because the arbitrary integration
constant $t_0$ was chosen to be
\begin{equation}
\label{e34} t_0 =
\frac{\beta_0}{2\,\alpha_0}=\frac{1}{2\,B_0}\,\sqrt{\frac{3\,\,\mu_0\,c^2(1-8\,\mu_0\,\omega
B_0^2)} {4\,\pi\,G}}\;.
\end{equation}
The time behaviour of the magnetic field is readily obtained from
(\ref{e27}) and (\ref{e31}). One finds
\begin{equation}
\label{e35}
B(t)=\frac{B_0}{(4\,\alpha_0^2\,t^2\,+\,4\,\alpha_0\,\gamma_0\,t\,+\,\delta_0)^{1/2}}\;,
\end{equation}
with the energy density and pressure defined by Eqs. (\ref{e17}),
(\ref{e18}), respectively. Note also that the Hubble parameter can
be written as
\begin{equation}
\label{e36} H=\frac{\dot{a}}{a}=\frac{\alpha_0\,\left
[2\alpha_0(t-t_0)+\beta_0\right]}{\left[4\alpha_0^2(t-t_0)^2+4\alpha_0\beta_0(t-t_0)+1\right]}
\;,
\end{equation}
which becomes for $t=t_0$
\begin{equation}
\label{e37} H_0\equiv H(t_0)
=\alpha_0\,\beta_0=B_0\,\sqrt{\frac{4\,\pi\,G\,(1-8\,\mu_0\,
\omega B_0^2)}{3\,\mu_0\,c^2}}\;.
\end{equation}
(In \cite{novello98a}, the notation $H$ is used to represent the
magnetic field.) From (\ref{e31}) we see that, for large $t$, we\
\ recover\ \ the\ \ classical\ \ solution\ \ for\ \ radiation \ \
dominated \ \ universes, $a(t)\propto t^{1/2}$. Alternatively, we
observe from the relation (\ref{e17}) that this solution is
recovered for values of t where the magnetic field obeys the
condition
\begin{equation}
\label{e38} 8\mu_0\omega B^2<<1.
\end{equation}
The most interesting feature of (\ref{e31}) is that the quadratic
function inside the parenthesis does not have real roots
$\omega>0$, being positive for any $t$. Therefore, the model is
non-singular with $a(t)$ reaching the minimum value $a_{min} =
a_0\,(8\,\mu_0\,\omega B_0^2)^{1/4}$ at a time
$t_{min}=-\frac{\gamma_0}{2\alpha_0}=t_0-\frac{\beta_0}{2\alpha_0}$.
It thus follows that the universe is a bouncing one: it begins
arbitrarily large at $t\ll t_{min}$, decreases until its minimal
value at $t_{min}$ and then begins to expand. The values of the
magnetic field and energy density at $t_{min}$ are

\begin{equation}
\label{e41} B(t_{min}) = \frac{1}{2\sqrt{2\mu_0\omega}},
\end{equation}

\begin{equation}
\label{e42} \rho(t_{min}) = 0 .
\end{equation}

Before proceed further, it is worth notice that if (\ref{e31})
describe rightly the evolution of the universe in the distant
past, it implies the existence of an inflationary era ($\ddot{a}
> 0$) on the interval
\begin{equation}
\label{e43} t_{min}-t_I\, <\, t\, <\, t_{min} + t_I\;,
\end{equation}
where
\begin{equation}
\label{e44a} t_I=\sqrt{\frac{3\,\mu_0^2\omega c^2}{\pi G}}.
\end{equation}

Figure 1 shows the scale factor, the magnetic field, the energy
density and the pressure as a function of time for a definite
value of $B_0$. From relations (\ref{e35}) and (\ref{e38}) we
stress that the classical solution is recovered for times much
larger than $t_{min}$ with the universe entering in the standard
radiation phase. As shown in Fig.1, the nonlinear corrections are
relevant only for $8\mu_0\omega B^2\, \gtrsim \,1/10$ or
$t\,\lesssim t_{min}+3\sqrt{8\mu_0\omega B_0^2}/2\alpha_0$.

\section{Nonsingular Models for
$\Lambda\neq 0$ } \label{s4}
For constant $\Lambda$, it is easy to see that equation
(\ref{efe1}) becomes
\begin{equation}
\label{efe3}
a\ddot{a}+{\dot{a}}^2-\frac{2}{3}\Lambda c^2 a^2- \left(\frac{64\pi
GwB_0^4a_0^8}{3c^2}\right)a^{-6}=0.
\end{equation}

Let us now search for an exact description in the presence of a
cosmological constant. By combining equations (\ref{e25}) and
(\ref{e27}) for a constant $\Lambda$ we find
\begin{equation}
\label{e45} \dot{Z}^2=16\,\left[\lambda Z^2+\alpha_0^2\,(Z-8\mu_0
\omega B_0^2)\right ]\;,
\end{equation}
where the auxiliary scale factor $Z$ and $\lambda$-term are
defined by
\begin{equation}
\label{e46} Z\equiv\left (\frac{a}{a_0}\right)^4\;,
\end{equation}
\begin{equation}
\label{e47} \lambda\equiv\frac{\Lambda c^2}{3}\;.
\end{equation}
Equation (\ref{e45}) can be easily integrated to give
\begin{equation}
\label{e48} a(t) = a_0 \left(\frac{1}{4\lambda}\right
)^{1/4}\,\left
[C_0\,e^{4\sqrt{\lambda}\,(t-t_0)}+\frac{D_0}{C_0}\,e^{-4\sqrt{\lambda}\,(t-t_0)}-2\,
\alpha_0^2\right ]^{1/4}\;,
\end{equation}
where
\begin{equation}
\label{e49} C_0\equiv
\alpha_0^2+2\lambda+2\sqrt{\lambda\,(\lambda+\alpha_0^2-8\,\alpha_0^2\,\mu_0\,\omega
B_0^2)}\;,
\end{equation}
\begin{equation}
\label{e50} D_0\equiv
\alpha_0^2\,(\alpha_0^2+32\,\lambda\,\mu_0\,\omega B_0^2)\;.
\end{equation}

The magnetic field is
\begin{equation}
\label{e51} B(t) = 2B_0\sqrt{\lambda}\,\left [C_0\,
e^{4\sqrt{\lambda}\,(t-t_0)} +\frac{D_0}{C_0}\,
e^{-4\sqrt{\lambda}\,(t-t_0)}-2\,\alpha_0^2\right ]^{-1/2}\;.
\end{equation}

It is straightforward to see that the term inside the square
brackets of (\ref{e48}) is positive for all $t$ and that the scale
factor reaches its minimum value
\begin{equation}
\label{e52} a_{min} = a_0\,\left
[\frac{\alpha_0}{2\,\lambda}\,\left (
\sqrt{\alpha_0^2+32\,\lambda\,\mu_0\,\omega B_0^2}-\alpha_0\right
)\right ]^{1/4}
\end{equation}
at
\begin{equation}
\label{e53} t_{min}= t_0 + \frac{1}{8\sqrt{\lambda}}\,\ln\left
(\frac{D_0}{C_0^2}\right )\;.
\end{equation}
As in the previous case, the universe bounces at $t_{min}$ and, if
the solution would effectively hold near $t_{min}$, an
inflationary phase would take place for all values of $t$ such
that
\begin{equation}
\label{e54} C_0^2
x^4-8\alpha_0^2C_0x^3+14D_0x^2-8\alpha_0^2\frac{D_0}{C_0}x
+\frac{D_0^2}{C_0^2} \, >\, 0\;,
\end{equation}
where
\begin{equation}
\label{e55} x\equiv e^{4\sqrt{\lambda}\,(t-t_0)}\;.
\end{equation}
The magnetic field at $t_{min}$ is
\begin{equation}
\label{e56} B(t_{min}) = \left
[\frac{\Lambda\mu_0c^4}{2\,\pi\,G\,\left
(\sqrt{1+\frac{8\,\Lambda\,\mu_0^2\omega c^4}{\pi\,G}}-1\right)}\right
]^{1/2}\;.
\end{equation}
>From relations (\ref{e38}) and (\ref{e51}) the  de Sitter
classical solution is recovered for

\begin{equation}
\label{e57} t<<t_0+\frac{1}{8\sqrt{\lambda}}\
ln\left(\frac{\alpha_0^4}{C_0^2}\right)
\end{equation}

and

\begin{equation}
\label{e58} t>>t_0+\frac{1}{8\sqrt{\lambda}}\
ln\left[\frac{(\alpha_0^2+32\mu_0\omega
B_0^2\lambda)^2}{C_0^2}\right].
\end{equation}

Similarly to what happens with solution (\ref{e28}), the classical
solution is recovered for times much larger than $t_{min}$. In
Figures 2 and 3 we show the scale factor, the magnetic field, the
energy density and the pressure as a function of time, for some
values of $\omega$, $\Lambda$ and $B_0$. In analogy with the
solution (\ref{e31}), we have that the nonlinear corrections are
relevant only for $8\mu_0\omega B^2\, \gtrsim \,1/10$ or
$t\,\lesssim t_0 +
\frac{1}{8\sqrt{\lambda}}\ln\left[\frac{(\alpha_0^2+160\,\mu_0\,\omega
B_0^2\lambda)^2}{C_0^2}\right]+\frac{1}{8\sqrt{\lambda}}\ln\left[\left(1+\sqrt{1-\frac{\alpha_0^2}{\alpha_0^2+160\,\mu_0\,\omega\,B_0^2\,\lambda}}\right)^2
\right]$.

\section{Nonsingular models for a time-dependent
$\bf \Lambda$} \label{s5}

The possible cosmological consequences of a decaying vacuum energy
density, or $\Lambda(t)$ cosmologies are still under debate in the
recent literature [14-30]. Such models may also be described in
terms of a scalar field coupled to a fluid component. Another
important motivation is its connection with the cosmological
constant problem. In general grounds, one may expect that a
decaying vacuum energy must play an important role on the universe
evolution (mainly in the very early Universe) and, probably, more
interesting, it may indicate suggestive ways to solve the
$\Lambda$-problem, as for instance, by describing the effective
regimes that should be provided by fundamental physics. In the
majority of the papers dealing with a time-varying $\Lambda$, the
decaying law is defined {\em a priori}, i.e., in a
phenomenological way. The most commonly postulated decaying laws
are those in which $\Lambda(t)$ decreases as some power either of
the scale factor $a(t)$ or the Hubble parameter $H$ (see
\cite{overduin98} for a quick review). Some authors have also
considered scaling laws formed by a combination of both quantities
\cite{CLW92}. As remarked before, these proposals are in
accordance but do not explain the difference of more than 100
orders of magnitude between the cosmological constant value at the
beginning of universe (provided by particles physics) and its
actual value estimated from cosmology. In general, the EFE imply
that once $\Lambda(t)$ is given one may integrate them for
obtaining $B(t)$ and $a(t)$. Conversely, for a given dependence of
$B(t)$, a unique time dependence for $\Lambda(t)$ is readily fixed
by the field equations.

Let us first analyze phenomenological models with a cosmological
term defined by $\Lambda = {3\beta}{c^{-2}} H^2$, where $\beta$ is
a positive parameter smaller than unity\cite{CLW92,CAL02}. The
differential equation driving the scale factor is readily derived
by combining relations (\ref{e17}), (\ref{e18}), (\ref{e21}) and
(\ref{e22}). One finds
\begin{equation}
\label{EFE4}
\dot{H}+4(1-\beta)H^2-(1-\beta)\frac{\alpha_0^2}{1-\beta_*^2}\left[1-\sqrt{1-\frac{4(1-\beta_*^2)}{\alpha_0^2}H^2}\right],
\end{equation}
where $H=\dot{a}/{a}$. The constants $\alpha_0$ is defined by
relation (\ref{e29}) whereas $\beta_*$ is given by
\begin{equation}
\label{c1} \beta_*\equiv \pm\sqrt{1-8\,\mu_0\, \omega (1-\beta)
B_0^2}\;.
\end{equation}
Now, separating the variables in (\ref{EFE4}) one finds

\begin{equation}
\label{EFE6}
\int_{H_0}^{H}{\left\{\frac{\alpha_0^2}{1-\beta_*^2}\left[1-\sqrt{1-\frac{4(1-\beta_*^2)}{\alpha_0^2}H^2}\right]
-4H^2\right\}^{-1}}dH= (1-\beta)(t-t_0).
\end{equation}
A simple integration results
\begin{equation}
\label{H2} H=\frac{\dot{a}}{a}=\frac{\alpha_0 \,\left
[2\alpha_0(1-\beta)(t-t_0)+\beta_*\right]}{\left[4\alpha_0^2(1-\beta)^2(t-t_0)^2+4\alpha_0\beta_*(1-\beta)(t-t_0)+1\right]}
\,,
\end{equation}
and integrating again, we obtain for the scale factor
\begin{equation}
\label{a3} a(t)= a_0\,\left [
4\,\alpha_0^2\,(1-\beta)^2(t-t_0)^2\,+\,4\,\alpha_0(1-\beta)\,\beta_*
\,(t-t_0)\,+\,1\right]^{[1/4(1-\beta)]}\;.
\end{equation}
This solution is nonsingular for $w>0$, with $a(t)$ reaching its
minimal value, $a_{min} = a_0\,(8\,\mu_0\,\omega (1 -
\beta)B_0^2)^{1/4(1 - \beta)}$, at a time
$t_{min}=t_0-\frac{\beta_*}{2\alpha_0(1-\beta)}$. It thus follows that the
universe is a bouncing one. It begins arbitrarily large at $t\ll
t_{min}$, decreases until its minimal value at $t_{min}$ and then
begins the expansion phase. For completeness, the expression for
the magnetic field is
\begin{equation}
\label{campo} B(t)=\frac{B_0(1-\beta)^{1/2}}
{[4\alpha_0^2(1-\beta)^{2}(t-t_0)^2+4\alpha_0(1-\beta)\beta_*(t-t_0)+1]^{1/2}},
\end{equation}
and at $t_{min}$, it is readily checked that $B(t_{min}) =
\frac{1}{2\sqrt{2\mu_0\omega}}$ and $\rho(t_{min}) = 0$. As one
should expect, in the limit $\beta=0$, all the results above for a
time dependent $\Lambda$-term reduce to that ones of $\Lambda=0$
(see equations (\ref{e27})-(\ref{e33})). Before proceed further,
it is worth notice the existence of an inflationary era ($\ddot{a}
> 0$) which depends on the value of the $\beta$ parameter. For
$1/2 < \beta \leq 1$ the universe always evolves trough an
accelerated expansion state. However, if $0 <\beta < 1/2$, it
inflates on the interval $t_{min}-t_I\, <\, t\, <\, t_{min} +
t_I$, where

\begin{equation}
\label{inf} t_I=\sqrt{\frac{3\,\mu_0^2\omega(1-\beta)^2 c^2}{\pi G(1-2\beta)}}.
\end{equation}

In Figures 4 and 5 we show the time dependence of the scale
factor, $\Lambda$-term, magnetic field, energy densities and
pressure for some selected values of $B_0$ and $\beta$. The main
conclusion is that the singularity must be avoided for a generic
time dependent $\Lambda$. {As shown in Figs. 4 and 5, the
nonlinear corrections are relevant only for ${8\mu_0\omega B^2\,
\gtrsim \,1/10}$  or
 ${t\,\lesssim t_{min}+3\sqrt{8\mu_0\omega(1-\beta)
 B_0^2}/2\alpha_0(1-\beta)}$.}

At this point one may ask by the inverse treatment, i.e., if the
singularity is avoided for a given $B(t)$. Such a question is
immediately answered by examining the simplest case, namely, that
one for which the magnetic field remains constant in the course of
the evolution. This possibility is clearly allowed by the
generalized energy conservation law (see (\ref{e26})). The energy
density of the magnetic field is kept constant because energy is
continuously drained from the decaying vacuum component to the
magnetic field. Actually, if $\Lambda$-term is maintained constant
the unique solution with a constant magnetic field is the trivial
one (${\bf B}=0$). Unlike the previous solutions, we note that
there is no an analogous classical solution for constant magnetic
field. Therefore, we will analyze such a possibility regardless of
any constraint on its domain of validity.

If $B(t)= B_0 = {\rm constant}$, the energy conservation law
yields
\begin{equation}
\label{e61} \Lambda(t)=\Lambda_0 + 3
K_0\ln\left(\frac{a}{a_0}\right )\;,
\end{equation}
where $\Lambda_0 \equiv \Lambda (t_0)$ and
\begin{equation}
\label{e63} K_0 \equiv
-\frac{16\,\pi\,G}{3c^4}\,\frac{B_0^2}{\mu_0}\,(1-16\,\mu_0\, \omega
B_0^2)\;.
\end{equation}

Substituting (\ref{e61}) into (\ref{e25}), we get for the scale
factor
\begin{eqnarray}
\label{e64} a(t)& = & a_0\, \exp\left
[\frac{K_0c^2}{4}\,(t-t_0)^2\,+\,H_0\,(t-t_0)\right ]\nonumber\\
 & = & a_0\, \exp\left(\frac{K_0c^2}{4}\,t^2+\beta_1\,
t+\beta_0\right )\;,
\end{eqnarray}
where
\begin{equation}
\label{e65} H_0 \equiv
\pm\sqrt{\frac{\Lambda_0c^2}{3}+\frac{4\,\pi\,G}{3c^2}\,\frac{B_0^2}{\mu_0}\,(1-8\,\mu_0\,
\omega B_0^2)},
\end{equation}
\begin{equation}
\label{e66} \beta_1\equiv -\left(\frac{K_0c^2}{2}\,t_0-H_0\right)
\,\,\,{\rm and}\,\,\,\beta_0\equiv t_0\,\left(\frac{K_0c^2}{4}\,t_0-H_0\right)
.
\end{equation}

The Hubble parameter is
\begin{eqnarray}
\label{e68} H(t)&=&\frac{K_0c^2}{2}\,(t-t_0) + H_0=\frac{K_0c^2}{2}\,t+\beta_1\;,
\end{eqnarray}
and we have $H=0$ for
\begin{equation}
\label{e69} t_c=
-\frac{2\beta_1}{K_0c^2}=t_0-\frac{2H_0}{K_0c^2}\;.
\end{equation}
At this point, the scale factor reaches the value
\begin{equation}
\label{e70} a(t_c)=a_0\,e^{-H_0^2/K_0c^2}\;.
\end{equation}

The behaviour of the solution will depend on the sign of the
constant $K_0$ (for $K_0 = 0$ we get the de Sitter solution).
 For ${K_0 >0}$ , that is, for ${B_0 >
1/(4\sqrt{\mu_0\omega})}$, the universe is always accelerated
(${\ddot{a}>0}$) e has a minimum size at ${t_c}$.

 A much more interesting solution is the one corresponding to
$K_0<0$ ($B_0 <1/(4\sqrt{\mu_0\omega})$). {For this range of
${B_0}$, ${a(t)}$, ${\ddot{a} > 0}$ for ${t< t_c-\sqrt{-2/ K_0
c^2}}$ and ${t> t_c+\sqrt{-2/ K_0c^2}}$ and ${a(t)}$ has a maximum
at ${t_c}$.} It is worth notice that the time interval $\Delta
t_{(NI)}$, prior to $t_c$, for which the solution is {\em not}
inflationary depends on the value of $B_0$ as
\begin{equation}
\label{e73} \Delta t_{(NI)} = 2\sqrt{-\frac{2}{K_0\,c^2}}=
\frac{2}{B_0}\,\sqrt{\frac{3\mu_0c^2}{8\,\pi\,G\,(1-16\,\mu_0\,\omega
B_0^2)}}\;.
\end{equation}

The cosmological term $\Lambda$ dominates the dynamics of universe
for values of $t$ where
\begin{equation}
\label{e74} \Lambda c^4 >8\pi G\rho .
\end{equation}

For both models ($K_0>0$ and $K_0<0$), the condition
(\ref{e74}) is satisfied by $t<t_3$  and  $t>t_4$  where
\begin{equation}
\label{e76} t_3=\frac{-2H_0 c-2\sqrt{\frac{16\pi G\rho}{3}}}{K_0c^3}
\end{equation}
and
\begin{equation}
\label{e77} t_4=\frac{-2H_0 c+2\sqrt{\frac{16\pi G \rho}{3}}}{K_0c^3}.
\end{equation}

In Figure 6, we show the scale factor and the cosmological term as
a function of time for $K_0 > 0$ and some values of $B_0$ and
$\lambda_0 = \Lambda_0c^2/3$. The same quantities have also been
plotted for $K_0 < 0$ in Figure 7.

Naturally, if one expects any such model to properly describe the
evolution of the real universe, it would be advisable to take into
account other matter fields, such as ultrarelativistic matter,
scalar fields or dust. In \cite{novello98a}, it was demonstrated,
for the case $\Lambda=0$, $B=B(t)$, that the presence of
ultrarelativistic matter with an equation of state $p_{(ur)}
=\rho_{(ur)}/3$ would just amount for a reparametrization of the
constants $B_0$ and $\omega$.

At this point, we would like to stress the mathematical
consistence of the whole set of solutions derived in the present
work. In general, there are 3 unknown functions: the scale factor
$a$, the magnetic field $B$ and the cosmological term $\Lambda$
(constant or time-dependent). As one may check for each case, the
number of unknown functions and equations coincide, the unique
exception is related to models containing a variable $\Lambda(t)$
term for which the phenomenological law $\Lambda = 3\beta
c^{-2}H^2$, has been considered (see Refs. \cite{CLW92,jackson}).


\section{CONCLUSION}

\label{s6}


We have examined whether nonlinear corrections to the Maxwell
electrodynamics may avoid the cosmic singularity occurring in flat
FRW universes. In brief, the answer is positive. We show that by
discussing a large class of analytical cosmological models under
three different assumptions. In the first case, the cosmological
$\Lambda$-term is identically zero and the dynamics is driven by a
time dependent magnetic field. This class generalizes the
particular solution previously found by De Lorenci {\em et
al.}\cite{novello98a}, and confirms their statement concerning the
avoidance of the initial singularity. In principle, since the
solutions are  non-singular, they potentially  solve the horizon
problem. We have also examined if the basic features of such
models remain true if new ingredients are introduced in the matter
content. In this concern, models with a constant and time
dependent $\Lambda$-term were studied with some detail. Again, for
both cases, the universe is also non-singular, bouncing at a
critical time when the scale factor reaches its minimum value.

For a decaying vacuum energy density we discuss two different
scenarios. In the first one, it was phenomenologically described
by $\Lambda(t) \sim H^{2}$ as assumed by several authors
\cite{CLW92,overduin98}. These models are nonsingular and resemble
the solutions with no $\Lambda$. The second scenario is a rather
curious solution which describes a universe driven by a constant
magnetic field. The time behaviour of the cosmological term is now
uniquely determined by the EFE as a logarithm of the scale factor.
It should be interesting to examine if such results are maintained
in the presence of other matter fields, as well as for universes
with non-zero curvature.

Finally, in analogy with the cosmological case, one may ask if
nonlinear terms in the Maxwell Lagrangian may remove the physical
singularity present in a charged black hole (Reissner-Nordstrom
solution). This problem will be discussed in a forthcoming
communication.

\acknowledgments The authors would like to thank an anonymous
referee whose comments improved the presentation of the paper. We
are also grateful to the CNPq (Brazilian Research Agency)  for
partial financial support.

\newpage
\begin{figure}
\vspace{.2in}
\centerline{\psfig{file=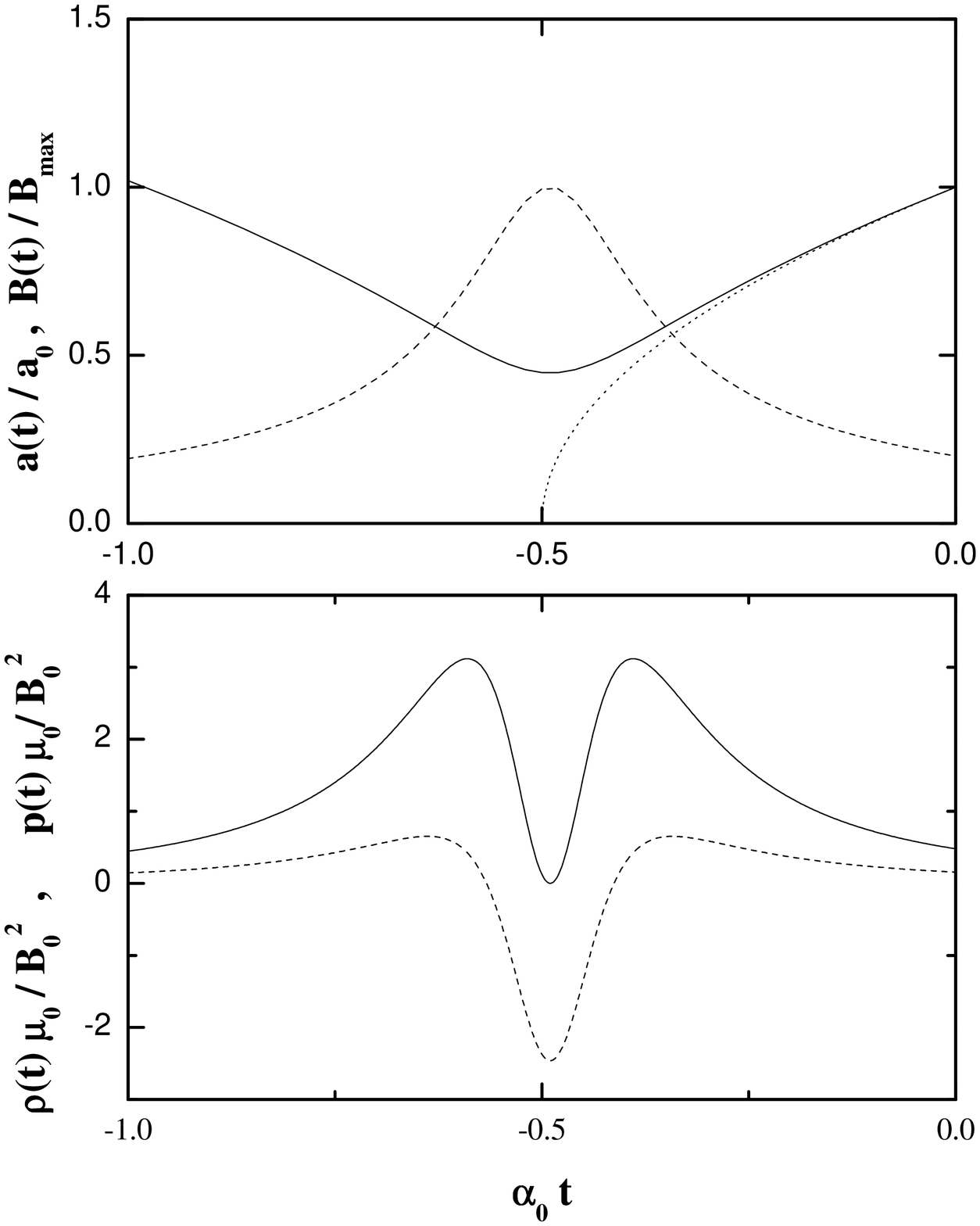,figure=Fig1c.eps,width=3.5truein,height=3.5truein}}
\caption{\protect{The upper panel shows the scale factor (solid
line), the magnetic field (dashed line) and the classical solution
(doted line) for $(\omega=0)$. The lower panel shows the energy
density (solid line) and the pressure (dashed line) for the model
with $\Lambda =0$ and $2B_0\sqrt{2\omega\mu_0}=0.2$. }}
\end{figure}
\begin{figure}
\vspace{.2in}
\centerline{\psfig{file=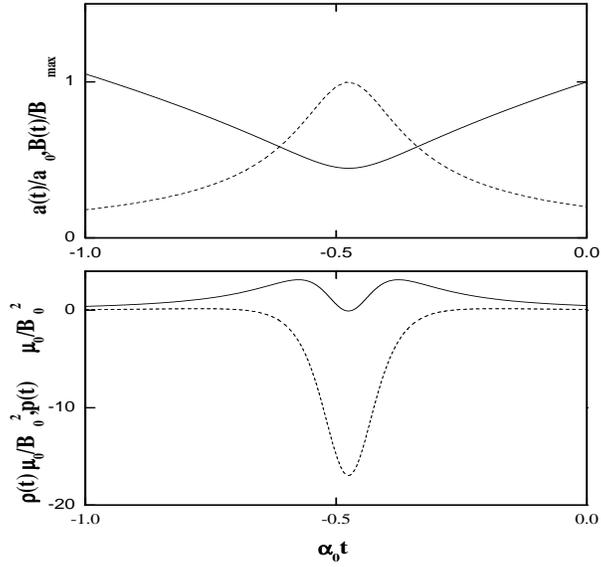,figure=Fig2c.eps,width=3.5truein,height=3.5truein}}
\caption{\protect{The upper panel shows the scale factor (solid
line) and the magnetic field (dashed line). The lower panel shows
the energy density (solid line) and
 the pressure (dashed line) for
the model with a constant non-vanishing $\Lambda$ . The
values for $\Lambda$ and $B_0$ are such that
$\sqrt{\lambda}/\alpha_0 = 0.4$ and
$2B_0\sqrt{2\omega\mu_0}=0.2$.}}
\end{figure}
\begin{figure}
\vspace{.2in}
\centerline{\psfig{file=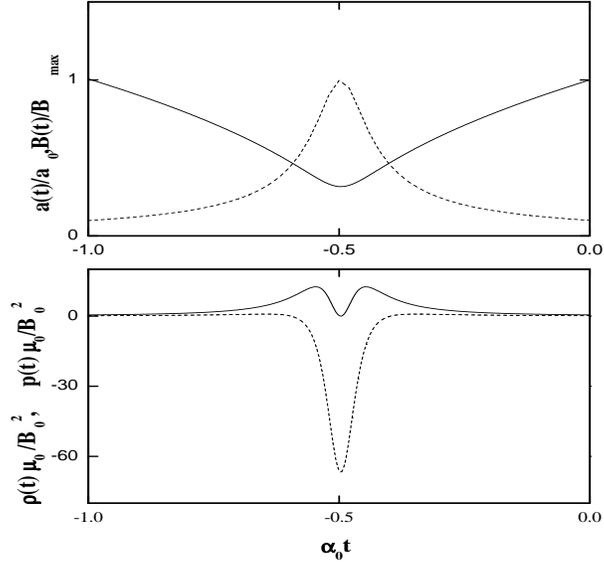,figure=Fig3c.eps,width=3.5truein,height=3.5truein}}
\caption{\protect{As in Figure 2 but for
$\sqrt{\lambda}/\alpha_0=0.01$ and
$2B_0\sqrt{2\omega\mu_0}=0.1$.}}
\end{figure}
\begin{figure}
\vspace{.2in}
\centerline{\psfig{file=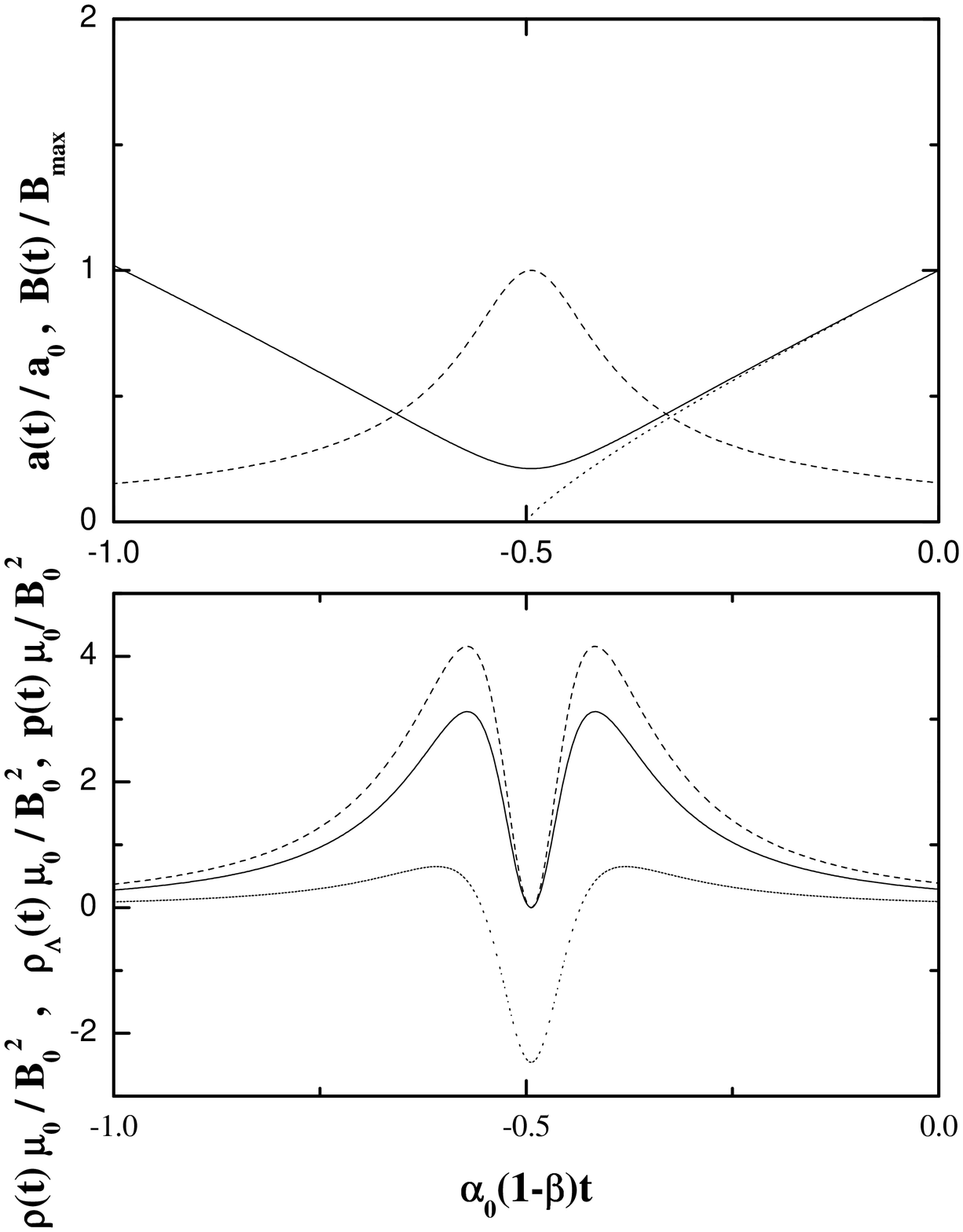,figure=cosmo1c.eps,width=3.5truein,height=3.5truein}}
\caption{\protect{The upper panel shows the scale factor (solid
line),  the magnetic field (dashed line) and the classical
solution (doted line) for $(\omega=0)$.  The lower panel shows the
energy density of magnetic field (solid line), the energy density
of the $\Lambda$ term (dashed line) and the pressure (doted line)
for the model with
 $\Lambda =\frac{3\beta}{c^2}H^2$, $\beta=0.4$ and
$2B_0\sqrt{2\omega\mu_0}=0.2$. }}
\end{figure}


\begin{figure}
\vspace{.2in}
\centerline{\psfig{file=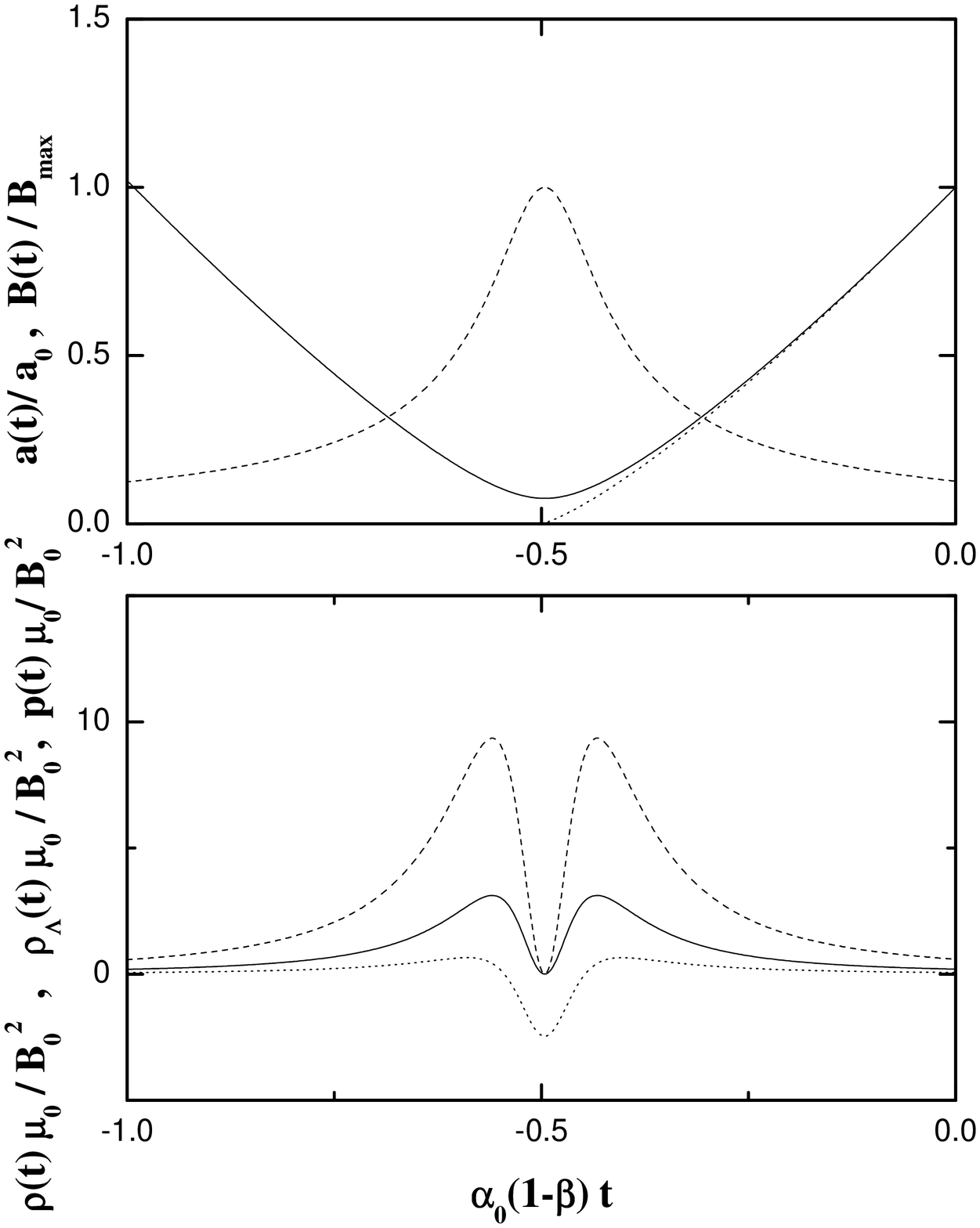,figure=cosmo2c.eps,width=3.5truein,height=3.5truein}}
\caption{\protect{The upper panel shows the scale factor (solid
line), the magnetic field (dashed line) and the classical solution
(doted line) for $(\omega=0)$. The lower panel shows the energy
density of magnetic field (solid line), the energy density of the
$\Lambda$ term (dashed line) and the pressure (doted line) for the
model with
 $\Lambda =\frac{3\beta}{c^2}H^2$, $\beta=0.6$ and
$2B_0\sqrt{2\omega\mu_0}=0.2$. }}
\end{figure}
\begin{figure}
\vspace{.2in}
\centerline{\psfig{file=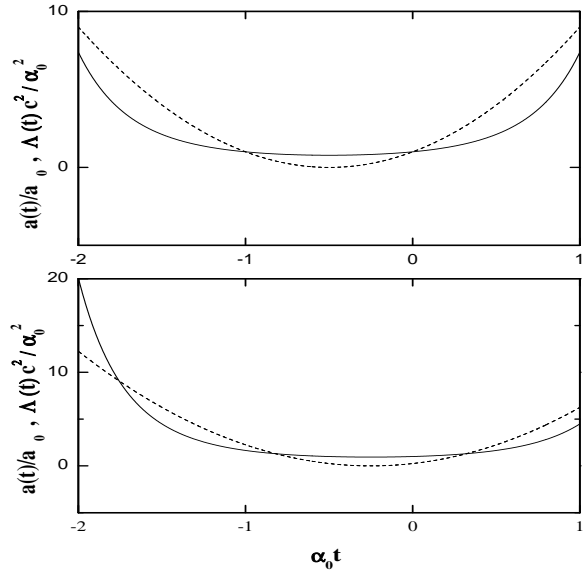,figure=Fig4c.eps,width=3.5truein,height=3.5truein}}
\caption{\protect{The scale factor (solid line) and the
cosmological term (dashed line) for the model with constant
magnetic field, time-dependent $\Lambda$, $K_0 > 0$
($2B_0\sqrt{2\omega\mu_0}=1$). In the upper panel
$\sqrt{\lambda_0}/\alpha_0 =1$ and the lower panel is for
$\sqrt{\lambda_0}/\alpha_0 =0.5$.}}
\end{figure}
\begin{figure}
\vspace{.2in}
\centerline{\psfig{file=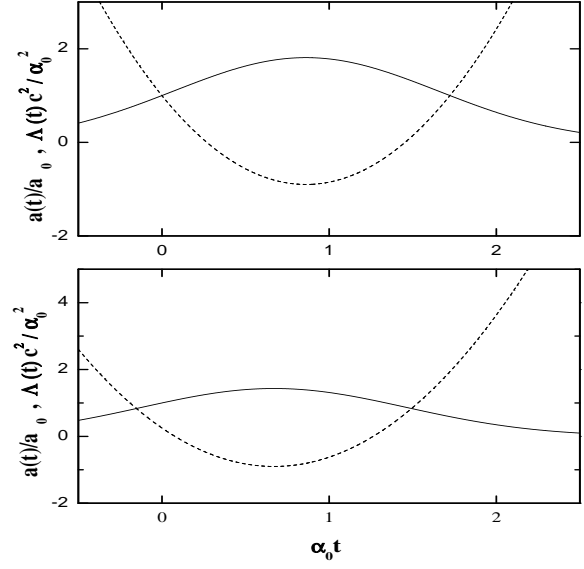,figure=Fig5c.eps,width=3.5truein,height=3.5truein}}
\caption{\protect{As in Figure 4 but for $K_0 < 0$
($2B_0\sqrt{2\omega\mu_0}=0.1$). In the upper panel
$\sqrt{\lambda_0}/\alpha_0 =1$ and the lower panel is for
$\sqrt{\lambda_0}/\alpha_0 =0.5$.}}
\end{figure}

\end{document}